\documentclass{optica-article}

\journal{opticajournal} 

\articletype{Research Article}

\usepackage{lineno}

\usepackage{graphicx}
\usepackage{subfigure} 
\usepackage{xcolor} 
\usepackage{nicefrac}

\usepackage{dcolumn}
\usepackage{bm}

\usepackage[utf8]{inputenc}
\usepackage[T1]{fontenc}
\usepackage{etoolbox}

\begin{document}

\title{Compact, open, and tunable two-dimensional Fabry-Pérot cavity}

\author{
Bruno Bender,\authormark{1} 
Andrea Bergschneider,\authormark{2}
Marcel Baer,\authormark{3} 
Martin Kroner,\authormark{4}
and Wolf Wüster\authormark{1,*}
}

\address{
\authormark{1}Institute of Applied Mathematics and Physics, Zurich University of Applied Sciences (ZHAW), 8401 Winterthur, Switzerland.\\
\authormark{2}Physikalisches Institut, University of Bonn, 53115 Bonn, Germany.\\
\authormark{3}Physics Department, ETH Zürich, 8093 Zurich, Switzerland.\\
\authormark{4}Institute for Quantum Electronics, ETH Zürich, 8093 Zurich, Switzerland.\\
}

\email{\authormark{*}wuew@zhaw.ch} 


\begin{abstract*} 
In order to enhance light-matter interactions, optically active material, like a semiconductor, can be embedded into an optical cavity. For small cavity mode volumes, which are required for reaching the strong coupling regime, this is typically done by fabrication of microcavities. While guaranteeing a mechanically stable cavity, these systems lack tunability and are difficult to implement for small samples like van der Waals heterostructures (vdW). Here we present a design for a fully tunable two-dimensional (2D) Fabry-Pérot (FP) cavity which can host any material. The cavity can be freely positioned between two confocal high-NA aspheric lenses which allows for optical transmission microscopy and spectroscopy in real and momentum space. In order to maintain maximal mechanical stability, the cavity mirrors are held in a monolithic titanium frame. A pre-loaded flexure joint allows for manual coarse tuning of the cavity length and mirror parallelism, while three piezo actuators in a tripod layout can be used for fine tuning. We find cavity length fluctuations smaller than 100 pm, measure a finesse of 360 and reach the lowest accessible $m=3$ cavity mode (one electric field antinode within the air gap) at a minimal physical mirror separation of 380 nm. In order to demonstrate the 2D nature of the cavity mode we perform momentum space imaging in transmission as well as spectroscopy to measure the cavity dispersion.
\end{abstract*}

\section{\label{sec:level1} Introduction}
Coherent superposition between light and matter excitations, which can be achieved by embedding a direct-bandgap semiconductor in a high quality optical cavity\cite{Shang2023}, can lead to emergent quantum phases with novel physical properties. The most prominent example for this was the demonstration of Bose-Einstein condensation of exciton polaritons, a coherent superposition of cavity photons and excitons in a two-dimensional (2D) semiconductor quantum well (QW)\cite{Kasprzak2006}.
This landmark experiment opened the field for the study of quantum fluids of light\cite{carusotto2013} with the discovery of superfluidity of polaritons\cite{Amo2009}, vortex formation\cite{Nardin2011}, or polariton lasing\cite{Schneider2013}.
Most of these remarkable achievements were obtained from QWs in monolithic, 2D Fabry-Pérot cavities. These devices are typically fully grown by molecular beam epitaxy (MBE) and hence the position of the QW with respect to the mirrors, as well as the mirror separation is defined upon the growth of the device. The anisotropic growth of the crystal that is inherent to this method, then allows for sufficient tunability of the detuning between the cavity mode and the QW exciton resonance energy by moving the optical excitation/detection spot along the growth gradient\cite{Chervy2020} and thereby changing the cavity resonance frequency. While this approach leads to a mechanically perfectly stable cavity there are a few shortcomings. In particular, this can only be used for devices where the optically active material can be grown in the same process as the distributed Bragg reflectors (DBR) serving as mirrors.
Recently many new materials with interesting optical properties emerged which cannot easily be integrated in such a cavity. These are 2D van der Waals materials such as transition metal dichalcogenides (TMD)\cite{Mak2016} or perovskites\cite{Manser2016}. Especially for vdW heterostructures, cavity integration is challenging due to their small lateral size of only a few 10µm. So far, two approaches have been successfully implemented. A monolithic 2D cavity can be formed by either placing a microscopic "flake" of DBR coating on top of a vdW heterostructure\cite{Rupprecht2021}. Alternatively, a mirror can be grown on top of the heterostructure \cite{Lopriore2025, Knopf2019} by i.e. plasma-enhanced chemical vapour deposition. Despite its challenging fabrication effort, this yielded a monolithic 2D cavities with a high quality factor (approaching 4000 in the first case and between 4000 and 5000 in the second, limited by absorption in the graphene gates) due to their inherent mechanical stability. However, these devices lack a direct degree of freedom to tune the cavity-exciton detuning in situ. To overcome this limitation, open, tunable zero-dimensional (0D) cavities \cite{Hunger_2010, Fisicaro2024, hoang2026} were successfully applied to TMD materials\cite{Dufferwiel2015, Sidler2017, Vadia2021, Drawer2023}. However, such 0D cavities do not offer access to in-plane modes and typically achieve high finesse only at the cost of limited optical access and restricted mode geometries. These 0D cavities are typically formed by one curved and one flat mirror in a hemispherical configuration which guarantees a stable optical cavity mode and hence rendering the device rather insensitive to small angles between the two mirrors. Therefore, only the mirror distance needs to be controlled in situ to tune the cavity modes. Nevertheless, angle control using slip-stick operated goniometers has been applied to arrays of 0D cavities in order to gain control of the resonance energies of the individual cavities \cite{Lackner2025}.
Most open, tunable 2D cavities that have been implemented so far are based on two parallel mirrors held in a monolithic assembly with one of the two mirrors being mounted on a piezoelectric actuator or a slip-stick positioner that controls the mirror distance \cite{Krol2020, Krol2023, Lackner2021, Lackner2025}. Parallelism between the two mirrors in these devices can usually not be controlled in-situ but is achieved to some extend by bringing the two mirrors into contact. This typically limits the tunability, and a finite residual angle between the mirrors limits the finesse of the cavity,~\cite{Krol2020, Lackner2025}. 
Further, this approach benefits from large contact areas between the mirrors, which makes electrical contact to the sample difficult. Finally, all of the previously demonstrated open, tunable 2D cavity devices did not allow for transmission spectroscopy. While for photoluminescence spectroscopy this is usually not an issue, resonant optical excitation is best performed in a transmission geometry, since specular reflection from the cavity can be avoided. In particular for photon correlation spectroscopy \cite{Levinsen2019} it is very beneficial when all the transmitted photons can be analyzed and no further filtering of the excitation light is necessary, as it has been reported on 0D cavities \cite{Delteil2019 ,Scarpelli2024}. 
Here, we present a compact, open and tunable 2D FP cavity with a high mechanical stability. It allows for angle-resolved transmission spectroscopy, positioning of the cavity mode, and integration of arbitrary (2D) materials. While maintaining a sufficiently compact size that is compatible with cryogenic operation, it features in situ control of mirror distance and wedge. Having control over the wedge of the FP cavity allows for experiments in which polariton propagation can be tuned. The layout is based on a monolithic assembly of a rigid mount holding the two planar mirrors which allows for pre-alignment with coarse tuning screws and with piezo actuators for in situ fine tuning of the mirror distance and relative angle between the mirrors. We characterize the cavity setup at room temperature and find a Q-factor of 3750 and a finesse of 360. With a minimal physical mirror separation of approximately 380 nm we achieve mode volumes that are compatible with the strong coupling condition for typical vdW exciton transitions. 

\newpage
\section{\label{sec:level1} Cavity Design and Optical Setup}
\begin{figure}[h!]
    \includegraphics[width=\linewidth]{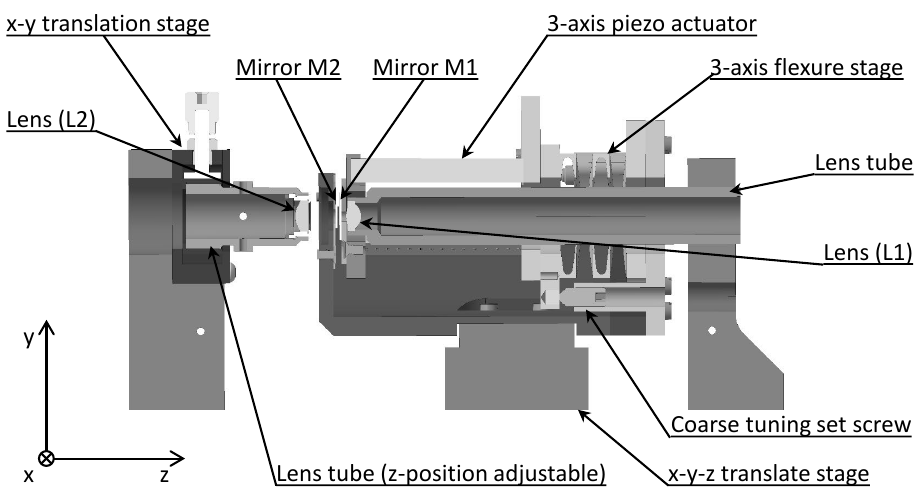}
\caption{\label{fig:cavity_assembly} Sectional view of the cavity assembly: The piezo assembly with the mirror holder is shown in the center. On the right side, the flexure mount with the set screws and the long lens tube can be seen. The three piezo actuators and the set screws for the flexure mount are arranges on a circle in a 120° layout with a 60° angle between them. Therefore, in the sectional view only one piezo and one set screw are visible. The cavity mirrors are placed between the two lenses on the left side of the center. The setup is designed such that different lenses can be mounted requiring different positioning of the lenses from the cavity, as seen in this schematic.}
\end{figure}

In the following we describe the mechanical assembly of the cavity, the cavity mirrors, and the optical setup we used for characterization. A schematic of the cavity layout and optical beam path is shown in Figure \ref{fig:cavity_assembly}.

The design goal of the mechanical assembly of the cavity is three-fold. i) It should allow for transmission spectroscopy which requires free-space optical access from both sides to the cavity mirrors. ii) The assembly should be compatible with operation at cryogenic temperatures and high magnetic fields, with sufficient in situ tunability of the relative mirror distance and angle. This shall also include the positioning of the cavity with respect to the in- and out-coupling lenses. iii) The assembly should be mechanically stable with eigenfrequencies in the high kHz regime to reduce vibrations of the mirrors.

In order to fulfill these requirements the mirrors are held in a monolithic assembly machined predominantly from titanium. One mirror is held at a fixed position, while the other one can be positioned using piezo-electric actuators for in situ position control and coarse tuning screws for prealignment. The piezo assembly is designed so that the three piezos are glued onto flexure joints on the base plate and the mirror holder in a triangular geometry. This avoids excessive straining of the piezos upon actuation. By controlling the three voltages applied to the piezo actuators, we can control the wedge angle of the mirror as well as its distance with respect to the fixed mirror. The piezo actuators (PI P-885.91) have a stroke of 38$\mu$m at room temperature and $\approx 4\mu$m at cryogenic temperatures of 4 Kelvin. In order to reduce the size of the piezo assembly, a three-axis flexure joint could be designed that uses lever arms to amplify the stroke of smaller piezos. The piezo assembly is screwed onto a flexure mount which is pre-stressed using three set screws with a pitch of 0.35mm per rotation. Using these three screws the cavity length and mirror angle can be prealigned to compensate for imperfections in the sample or mirror mounting and thickness. The eigenfrequencies of the unstressed flexure mount are around 600 Hz but are expected to be shifted to much higher frequencies upon the applications of stress using the micrometer screws. Precise estimates of the frequency shift is difficult to estimate as the applied stress is unknown and depends on the alignment of the setup. Our cavity length stability measurements confirm no sharp resonances up to 10 kHz. In order to at least partially compensate for different thermal expansion coefficients between the titanium frame, the piezos, set screws, and the mirror mounts, a copper spacer is integrated into the assembly.
The whole assembly has a coaxial design into which a 12 mm diameter lens tube can be inserted. In this way, the cavity can be positioned between two confocal lenses (Thorlabs A375-B, NA=0.3, $f$=7.5 mm) to allow for transmission measurements. The cavity assembly is designed to be mounted on cryo-compatible positioners, enabling precise alignment within the focal volume of the lenses.

\begin{figure}[htb!]
    \subfigure[]{\includegraphics[width=0.42\linewidth]{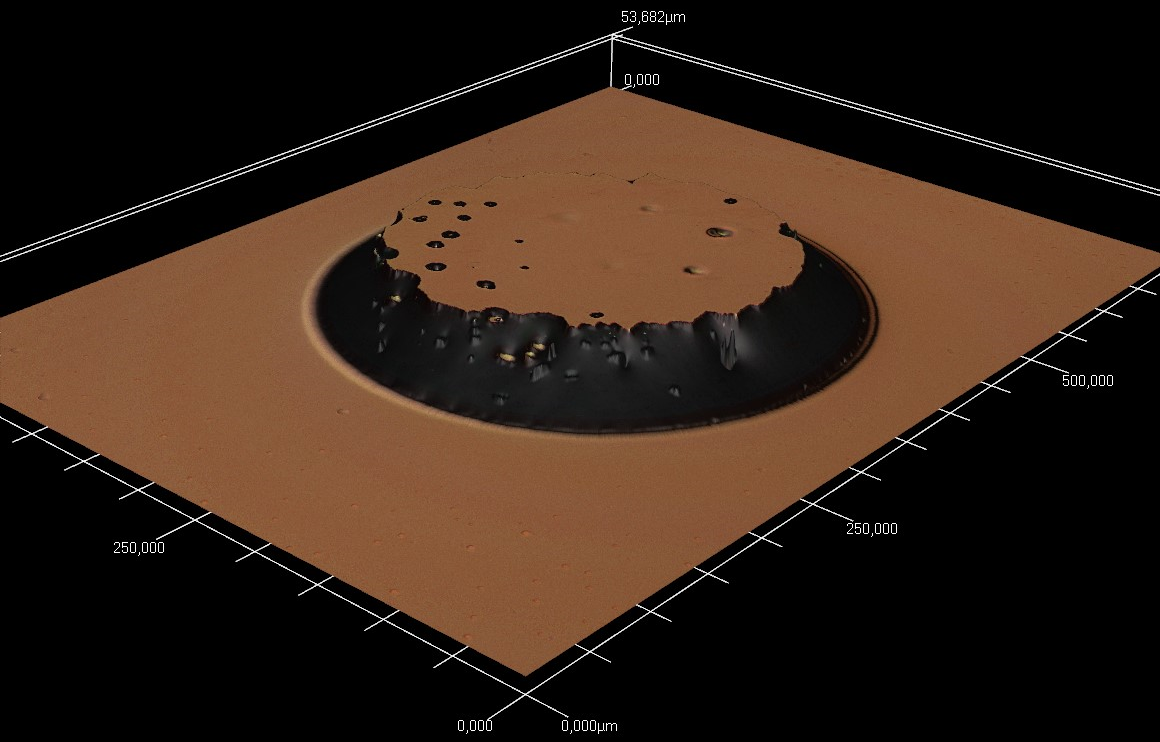}}
    \subfigure[]{\includegraphics[width=0.54\linewidth]{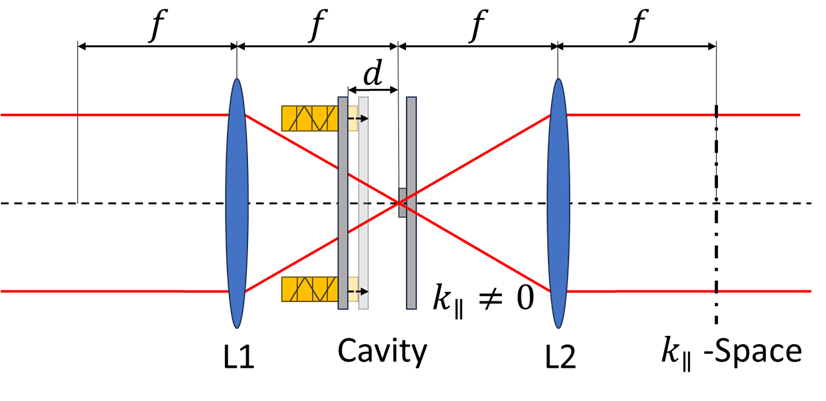}}
\caption{\label{fig:opt_setup_and_mesa} Panel (a) shows the mesa structure of the DBR mirror measured with the Keyence Laser Microscope VH-X1000. The diameter of the mesa is $d_{\textrm{mesa}} \approx 275 \,\mu \textrm{m}$ and the height is $h_{\textrm{mesa}} \approx 43 \,\mu \textrm{m}$. Panel (b) shows the optical setup for transmission spectroscopy, where collimated light enters lens L1 and is focused on the second cavity mirror. In this configuration, the cavity is excited with in-plane k-vectors greater than zero. The k-vectors are imaged on the back focal plane of lens L2.}
\end{figure}

The cavity is formed by two planar mirrors that consist of 500\,$\mu$m-thick fused silica substrates on which a total number of 24 layers of dielectric material alternating between Nb\textsubscript{2}O\textsubscript{5} and SiO\textsubscript{2} is deposited by Laseroptik GmbH (Garbsen, Germany) resulting in a nominal transmission of 500\,ppm at the DBR design wavelength of $\lambda_0 = 741$\,nm.
In order to reach cavity lengths on the order of a few wavelengths, one of the mirror substrates was processed with a mesa structure prior to the application of the coating. This allows the mirrors to come very close to each other without the edges colliding, which can occur due to the residual wedge, or non-planar nature of the mirrors. The mesa structure is fabricated using a chemical etching process. %
In a first step, the mesa structure was defined by a circular metal mask of $400\,\mu \textrm{m}$ diameter consisting of 30 nm chrome for enhanced adhesion, 400 nm of gold and few micrometers of hardened photoresist. Subsequently, the substrate was etched for 1 hour in hydrofluoric acid of $40\%$ concentration. The isotropic etching process occurred at a rate of about $0.7\,\mu$m/min resulting in a mesa height of around $40 \,\mu \textrm{m}$ and in underetching of the metal mask.
The 3D profile of this mesa structure, recorded by a laser microscope, is depicted in Figure \ref{fig:opt_setup_and_mesa} (a). It shows a round plateau with a diameter $d_{\textrm{mesa}} \approx 275 \,\mu \textrm{m}$ and a height of $h_{\textrm{mesa}} \approx 43 \,\mu \textrm{m}$. Small craters can be observed on the surface of the mesa, which result from insufficient protection of the mask during the etching process.
Figure \ref{fig:opt_setup_and_mesa} shows the optical setup for transmission microscopy in real and momentum space. The excitation light enters the first lens L1 and is focused into the cavity with a distribution of in-plane momenta according to the NA of the lens. 
The transmitted light is collected by the second lens L2. Due to the coincidence of the focal points of both lenses, each k-vector inside the cavity is focused on the back focal plane of lens L2 at a distinctive radial distance from the optical axis. By collecting light from the back focal plane, angle-resolved transmission spectra of the cavity can be measured. We used single mode optical fibers for both, delivering the excitation light as well as collecting the transmitted light. This alleviates the alignment of the coupling to the cavity as well as switching between different light sources and detectors. Integrating the fiber couplers directly into a rigid common frame together with the cavity mirrors could eliminate the remaining free-space alignment of the collection path, further improving the long-term mechanical stability of the setup.
We performed two different types of measurements, white-light transmission spectroscopy and resonant transmission microscopy:
In the first case we used two different excitation sources, a white-light laser (NKT Photonics SuperK Compact), and a superluminescent diode (Exalos EXS210025-01) providing different ranges of excitation wavelengths for the optical characterization. The spectra were recorded using a Princeton Instruments IsoPlane 160 spectrograph coupled to the ProEM 1024 camera.
For microscopy and cavity stability measurements, we excited the cavity with a laser diode (Thorlabs LP785-SF20) emitting multiple longitudinal modes, with a spectral width of approximately  1\,nm (see Appendix~C for the detailed spectrum). We imaged the transmitted mode in momentum space on a CMOS camera (Fujifilm X-T10).

\section{\label{sec:level1} Optical Characterization of the Cavity}
\begin{figure}[htb!]
    \includegraphics[width=\linewidth]{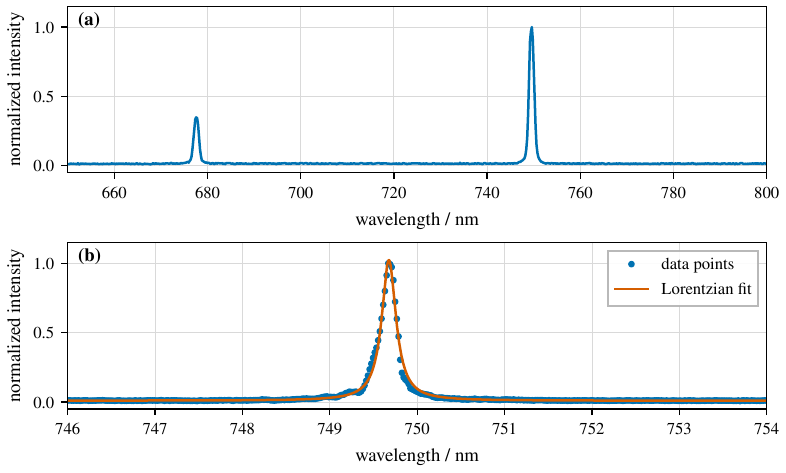}
\caption{\label{fig:Finesse_and_Q} (a) Transmission spectrum of cavity modes of order $m = 10$ and $m = 11$ inside the DBR stopband, separated by an FSR of $72.0 \pm 0.2$\,nm, recorded with an NKT SuperK white-light laser. (b) Detailed spectrum of the $m = 10$ mode recorded with an Exalos SLED, with a Lorentzian fit giving a FWHM of $\delta\lambda = 0.20 \pm 0.01$\,nm. The spectra were recorded with the IsoPlane 160 spectrometer and a ProEM1024 camera, using a 300\,grooves/mm grating and a 100\,$\mu$m slit in (a) and an 1800\,grooves/mm grating and a 25\,$\mu$m slit in (b). Since the two spectra were recorded with different light sources and settings, each is normalized to its maximum.}
\end{figure}

To characterize the performance of the cavity, we determined its finesse and quality factor by measuring the full width at half maximum (FWHM) of a resonance peak and the free spectral range (FSR) between adjacent modes. These values provide insight into the spectral resolution and the stored energy in the form of the electromagnetic field as well as the losses. The resonance condition for an FP cavity is given by Equation \ref{Eq:resonance_condition}:
\begin{equation}
    m \lambda = 2nd \cos\theta,
    \label{Eq:resonance_condition}
\end{equation}
where $m$ is the longitudinal mode number, $\lambda$ is the mode wavelength, $n$ is the refractive index inside the cavity, $d$ is the cavity length, and $\theta$ is the angle of propagation with respect to the cavity normal. We measure the cavity mode spectrum by collecting the transmitted light from the back focal plane of L2 behind the cavity. Figure \ref{fig:Finesse_and_Q}~(a) displays the measured cavity modes ($m=11$) at $\lambda_{m=11} = 677.59 \pm 0.2$\,nm and ($m=10$) at $\lambda_{m=10} = 749.54 \pm 0.2$\,nm, from which we determine an FSR of $72.0 \pm 0.2$\,nm (uncertainty: about one detector pixel). Since the simple model of Equation~\ref{Eq:resonance_condition} neglects the wavelength-dependent reflection phase of the DBR mirrors, we determined the mode numbers using a transfer-matrix calculation of the cavity based on the nominal layer design of the mirrors provided by the manufacturer (see Section~\ref{sec:tunability}). This yields the mode numbers stated above and confirms the mode numbers given in Figure \ref{fig:wedge_and_tunability}~(a). Figure \ref{fig:Finesse_and_Q}~(b) shows details of the resonant cavity mode (m=10) at 750 nm. We fitted the mode with a Lorentzian profile with constant background (free parameters: amplitude, center wavelength, FWHM and background) to all data points between 745 and 755\,nm, without prior background subtraction, and obtain a FWHM of $\delta\lambda = 0.20 \pm 0.01$\,nm. The statistical fit uncertainty is only 0.002\,nm; the quoted uncertainty additionally includes a systematic contribution estimated from the finite spectrometer resolution and the slight asymmetry of the resonance. Based on these values, we calculated a finesse of $\mathcal{F} = \mathrm{FSR}/\delta\lambda = 360 \pm 20$ and a quality factor of $Q = \lambda/\delta\lambda = 3750 \pm 190$, where the uncertainty is dominated by the linewidth. For comparison, state-of-the-art monolithic FP cavities report Q=3800 and a finesse of $\mathcal{F} = \frac{FSR}{\delta\lambda}$ = 2440 at a cavity length of $\approx 400$ nm \cite{Rupprecht2021}. Due to the finite transverse detection mode size of approximately 60 $\mu\mathrm{m}$ (determined by the finite diameter of the lens in the detection path), we collect a finite number of in-plane $k$-vectors. As a result, the measured quality factor is expected to be lower than the theoretical value. For an FP cavity consisting of two DBR mirrors with a reflectivity of R$\approx 99.95\%$, we expect a theoretical Q=65000 and $\mathcal{F} = 6300$ at a FSR=72nm. Due to k-averaging, we simulated that the maximum values that can be detected will be Q $\approx 10000$ and $\mathcal{F}\approx 930$. The experimentally measured values are roughly two to three times smaller. We believe that we are limited by defects and a residual wedge between the mirrors in the area of the $d_{\mathrm{detect}} = 60 \mu\mathrm{m}$ large transverse detection mode. In Figure \ref{fig:opt_setup_and_mesa} (a) the defects in the $d_{\mathrm{mesa}} \approx 275 \,\mu \textrm{m}$ large mesa are clearly visible.

We observed that the defect density can be substantially reduced through a thorough cleaning procedure prior to the application of the metallic etch mask, which could improve the achievable finesse. Independently, potential substrate curvature caused by strain from the dielectric coating could be reduced by using a thicker substrate or by applying a compensation layer to the backside of the substrate, improving the mirror parallelism. A further, related design parameter is the size of the detection mode, which can be adjusted by choosing a suitable combination of lenses in the relay optics while remaining fiber-mode-matched to the detection fiber. Enlarging the detection mode reduces the k-averaging that currently caps the finesse well below the mirror-limited value, so that a larger detection mode would in principle allow a higher finesse to be reached. This is not without limits, however: a larger detection mode necessarily samples a larger area of the mesa and can only be increased together with the mesa diameter itself, so that any residual defects or curvature within that area are correspondingly more likely to be sampled as well. Moreover, since the mesa exists specifically to keep the mirrors from touching at minimal separation, growing it beyond a certain size increases the risk of edge contact before the desired minimum mirror separation is reached. An optimal choice of mesa diameter and detection mode size therefore balances the resulting reduction in k-averaging against these competing effects, and benefits from being combined with the lower-defect, lower-curvature mirrors described above.

\newpage

\subsection{Length Tunability and Mirror Parallelism}\label{sec:tunability}
\begin{figure}[htb!]
    \subfigure[]{\includegraphics[width=0.48\linewidth]{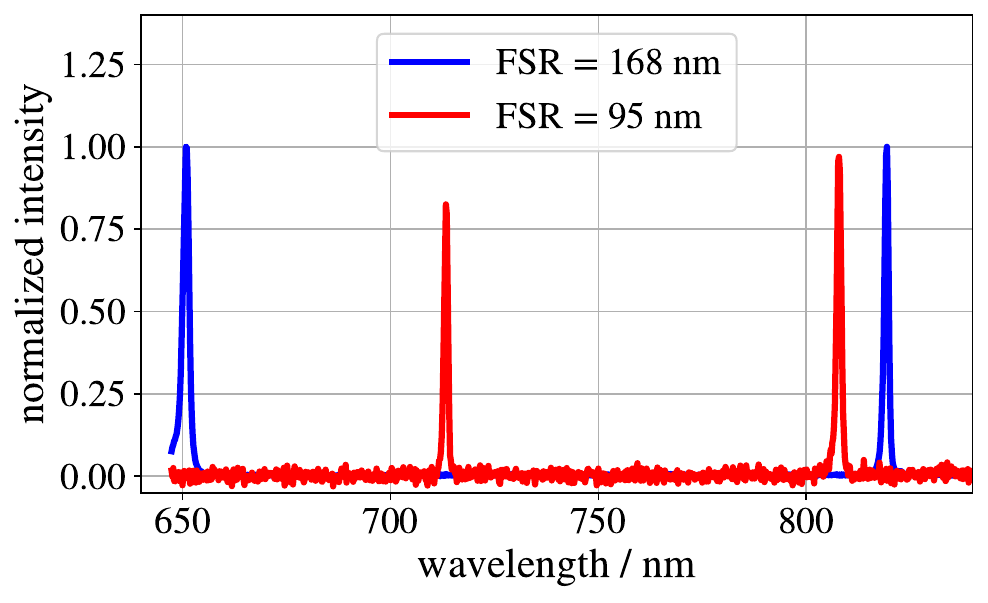}}
    \subfigure[]{\includegraphics[width=0.48\linewidth]{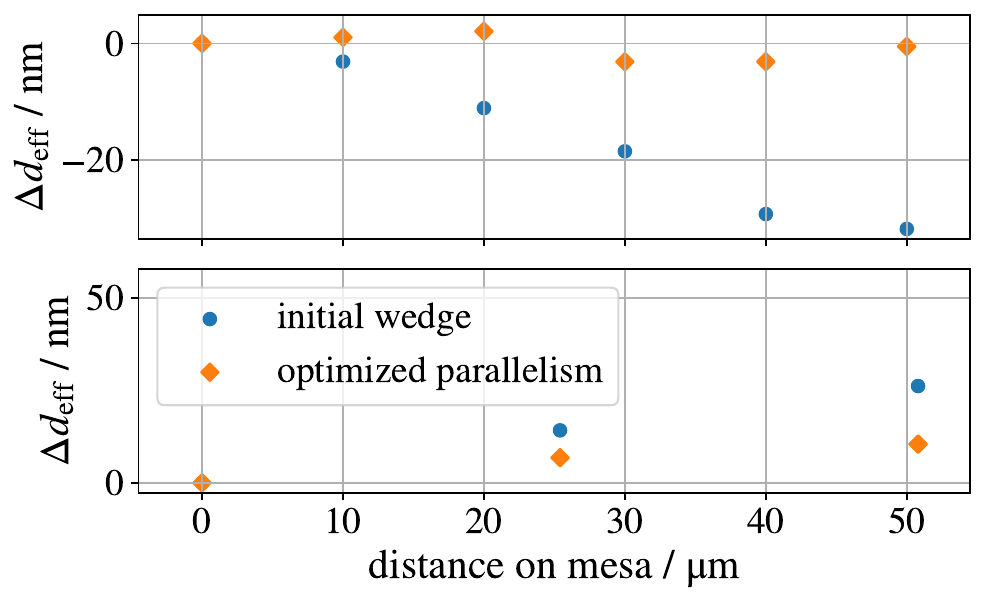}}
\caption{\label{fig:wedge_and_tunability} Tunability of the planar 2D cavity: (a) demonstrates the tunability of the cavity length by showing the free spectral range at different voltages applied to the piezo actuators. The red (blue) data show two longitudinal transmission modes at 713 nm (650 nm) with mode number m=9 (m=5) and at 808 nm (818 nm) with mode number m=8 (m=4). This corresponds to an effective cavity length of $d_{\mathrm{eff}} = 3.21\,\mu \mathrm{m}$ ($d_{\mathrm{eff}} = 1.64\,\mu \mathrm{m}$). (b) illustrates the optimization of mirror parallelism by scanning the cavity mode across the mesa and recording the spectral position of the resonance peak.}
\end{figure}

The tunability of the cavity is demonstrated by applying different voltages to the piezoelectric actuators and by determining the effective cavity length $d_{\textrm{eff}}$. For this purpose, we measured the free spectral range for actuator voltage sets of (52, 38, 43) V and (60.6, 46.6, 51.6) V. As before, the FSR was measured from the spacing between adjacent cavity mode peaks and converted to an effective cavity length using Equation \ref{Ey:FSR}:
\begin{equation}
    \mathrm{FSR} = \frac{\lambda_0^2}{2nd_{\mathrm{eff}} + \lambda_0}
    \label{Ey:FSR}
\end{equation}
The recorded transmission spectra are shown in Figure \ref{fig:wedge_and_tunability} (a). The FSR values of 95 nm and 168 nm correspond to cavity lengths of 3.21 µm and  1.64 µm, respectively. The blue data correspond to the minimum mirror separation, where two transmission modes still lie within the stopband of the DBR mirrors. We could further reduce the cavity length to an effective length of 1125 nm (mode number $m=3$ at 750\,nm). This corresponds to an approximate physical DBR mirror separation of 378\,nm. In contrast to a metallic mirror, the electric field of a DBR mirror penetrates into the mirror stack and decays exponentially. We performed a TMM calculation using the open-source Python package \texttt{tmm}~\cite{Byrnes2016} for the longitudinal mode $m=3$ tuned to the DBR design wavelength of 750\,nm and identified where the electric field drops to $1/e$ of its maximum value. We find a penetration depth of $L_{\mathrm{DBR}} = 216\,$nm per mirror and a mean effective refractive index of $n_{\mathrm{eff}} = 1.73$. The physical mirror separation $L_{\mathrm{cav}}$ follows from the effective optical length $L_{\mathrm{opt}} = n_{\mathrm{air}}L_{\mathrm{cav}} + 2n_{\mathrm{eff}} L_{\mathrm{DBR}}$, where $L_{\mathrm{opt}} = 1125\,$nm is determined from Equation~\ref{Eq:resonance_condition} and $n_{\mathrm{air}}=1$. Using the approximate formula $L_{\mathrm{DBR}} = \frac{\lambda_{\mathrm{DBR}}}{4 \Delta{n}} \approx 223\,$ nm we obtain similar results\cite{Brovelli1995}. We can achieve even smaller mode volumes (mode $m=3$ moved to shorter wavelengths), but we did not observe mode $m=2$ appearing (which would correspond to a physical mirror distance of only 3\,nm at 750\,nm center wavelength). These measurements confirm the ability to tune the cavity length and achieve the strong coupling condition $g = \frac{1}{4}\left(\hbar \gamma_{\text{c}} + \hbar \gamma_{\text{ex}}\right)$ for vdW excitons where we use a cavity photon linewidth of $\hbar \gamma_{\mathrm{c}} \approx 0.4\,\mathrm{meV}$, a typical vdW exciton linewidth of $\hbar \gamma_{\mathrm{ex}} \approx 1\,\mathrm{meV}$ and a coupling strength of $g= \sqrt{\frac{2\hbar \gamma_{\mathrm{c}}}{d_{\mathrm{eff}}}} = 20\,\mathrm{meV}$ \cite{Skolnick1998}. In monolithic cavities typical normal mode splittings of $2g = 31\,\mathrm{meV}$ are reported\cite{Rupprecht2021}.
To assess the parallelism of the cavity mirrors, we optically excited a spot on the mesa and monitored the spectral position of the corresponding cavity mode. The excitation spot was translated across the mesa surface by displacing the cavity using the linear stages, enabling scanning along two orthogonal directions. During this, the spectral position of the mode $\lambda_{m,d}$ was recorded as a function of the excitation spot position, allowing us to evaluate variations in the optical path length across the mesa surface and thus quantify the degree of mirror parallelism. From the spectral shift of the cavity mode $\Delta\lambda_m$ the difference in the effective cavity length $\Delta d_{\mathrm{eff}}$ is calculated according to Equation \ref{Eq:eff_cavitylength_from_mode_shift}:
\begin{equation}
\Delta\lambda_{m} = \lambda_{m, d1} - \lambda_{m, d2} = \frac{2n\cos(\theta)}{m} \Delta d_{\mathrm{eff}}
\label{Eq:eff_cavitylength_from_mode_shift}
\end{equation}
The results are shown in Figure \ref{fig:wedge_and_tunability} (b). We observed that in the initial state the optical path length shifted by 31.8 nm as the excitation spot was moved 50 $\mu$m in x-direction across the surface of the mesa, corresponding to a wedge angle of 637 $\mu\mathrm{rad}$. After optimizing the wedge of the mirror, the largest deviation in y-direction was reduced to 10.3 nm, corresponding to a residual wedge angle of 204 $\mu\mathrm{rad}$. In the x-direction the wedge could almost be eliminated. Typical wedge angles in MBE-grown monolithic cavities with growth gradient of monolithic cavities are on the order of 32 $\mu\mathrm{rad}$\cite{Wuester2015}.

\subsection{Cavity Dispersion}

The characteristic feature of two-dimensional FP cavities is the angular dispersion, which describes the shift of the cavity resonance with the in-plane momentum of the photon. To experimentally characterize this dispersion, we performed angle-resolved transmission spectroscopy using lenses with a numerical aperture (NA) of 0.3. The white light excitation was focused into the cavity using lens L1, and the transmitted light was collected from the back focal plane of lens L2 placed behind the cavity, which corresponds to the Fourier plane and allows momentum-resolved measurements.\\
We recorded momentum-resolved spectra by imaging the back-focal plane of the lens L2 in Figure \ref{fig:opt_setup_and_mesa} (b) onto the facet of a single mode fiber. To this end a third lens is introduced into to collection path ontop of L2 and a fiber coupling lens. In between these two lenses a mirror in a tilt stage can be used to control which position in the back focal plane of L2, and thereby which cavity mode with a specific in-plane momentum $k_{||}$ is coupled into the fiber. The transmission spectrum is then measured while scanning the tilt angle of the mirror using a stepper motor. For larger mirror tilts, the coupling efficiency of the transmitted light into the fiber decreases significantly. As a result, our measurements were limited to in-plane momenta up to $1.5\,\mu \text m^{-1}$. Figure \ref{fig:cavity_disperion} shows the measured transmission through the cavity as a function of cavity photon in-plane momentum and energy. The energy dispersion of a cavity is given by Equation \ref{Eq:energy_dispersion}:

\begin{figure}[htb!]
    \includegraphics[width=\linewidth]{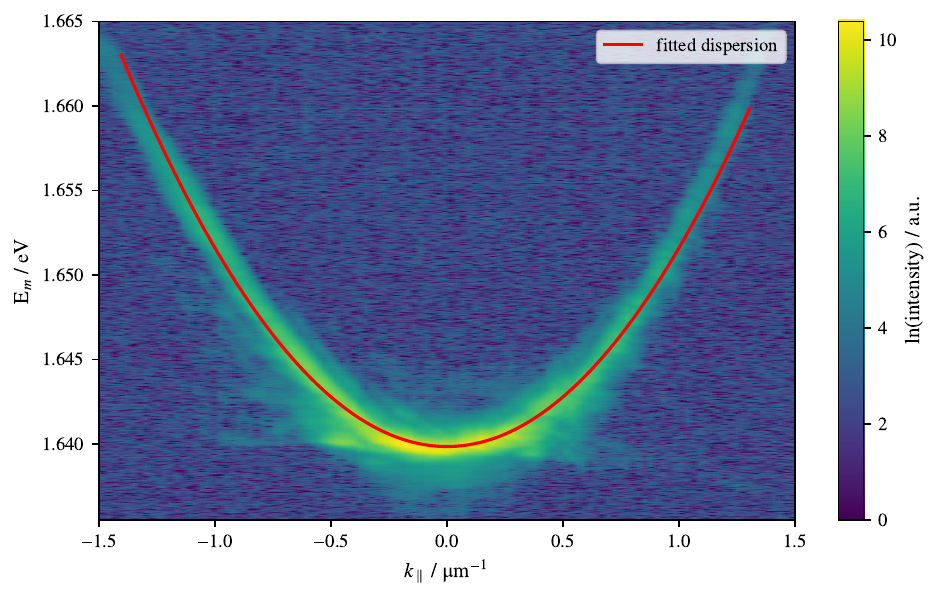}
\caption{\label{fig:cavity_disperion} Angle-resolved transmission spectrum of the open 2-dimensional cavity. The color map shows the logarithmic transmitted intensity as a function of in-plane momentum and photon energy, showing the parabolic dispersion of the cavity mode. The red line represents the theoretical fit to the dispersion.}
\end{figure}

\begin{equation}
E_{cav} = \hbar\frac{c}{n}\sqrt{k_\parallel^2 + k_\perp^2 }\approx E_0 + \frac{\hbar^2 k_\parallel^2}{2m_{cav}}
\label{Eq:energy_dispersion}
\end{equation}

We determined the peak position of the cavity mode in each angle-resolved spectrum from its local intensity maximum and fitted the dispersion relation according to Equation \ref{Eq:energy_dispersion} for $-1.4\,\mu\mathrm{m}^{-1} \le k_\parallel \le 1.3\,\mu\mathrm{m}^{-1}$, using $E_0$, the curvature (effective photon mass $m_{cav}$) and an angular offset of the scan as free parameters. We obtain $E_0 = 1.6399 \pm 0.0002$\,eV (fit uncertainty); within the fitted range, the data deviate from the parabola by 0.6\,meV (rms) over an energy range of 22\,meV. Since the angular scale of the steering mirror could not be determined precisely from the setup geometry, the $k_\parallel$-axis was calibrated using the cavity length obtained from the FSR measurement. The fitted curvature is therefore consistent with the cavity length by construction and not an independent result; the calibration only rescales the $k_\parallel$-axis and affects neither the parabolic shape of the dispersion nor $E_0$. The fitted dispersion is shown as the red line in Figure \ref{fig:cavity_disperion}. The measured data follow the expected quadratic behavior, consistent with two-dimensional photon confinement. However, at higher in-plane vectors $k_\parallel$, deviations from the ideal parabola become apparent, including an asymmetric shape of the dispersion curve.

\subsection{Fourier-Ring Imaging}

\begin{figure}[htb!]
    \subfigure[]{\includegraphics[width=0.46\linewidth]{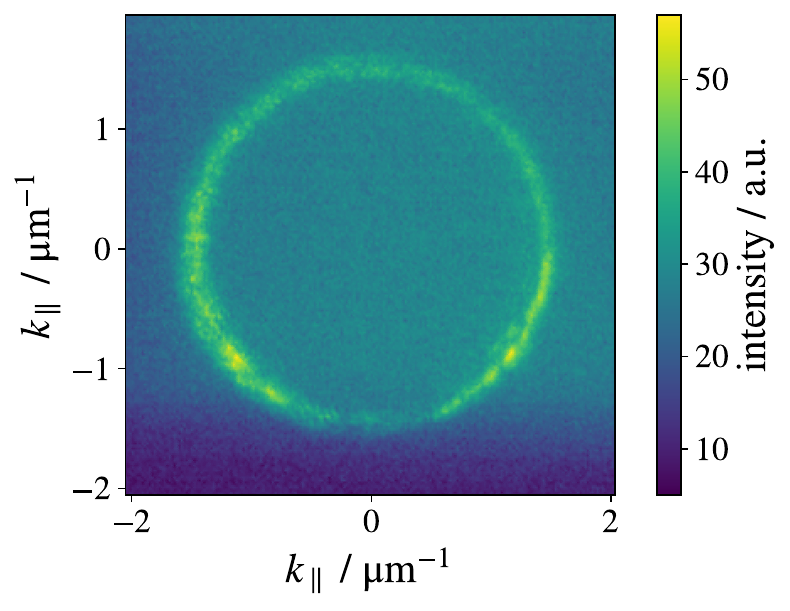}}
    \hspace{0.05\textwidth}  
    \subfigure[]{\includegraphics[width=0.46\linewidth]{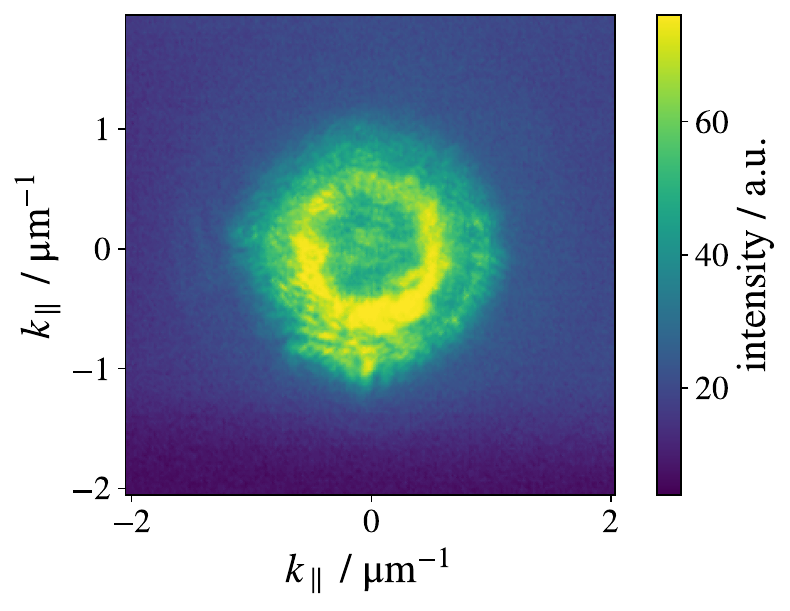}}
\caption{\label{fig:fourier_rings} Fourier-plane images of the transmission through the open 2D cavity. The left image shows a narrow Fourier ring that is highly symmetrical. The right image shows a ring with a smaller diameter that is much broader. The increased ring width at smaller in-plane momentum can be explained by the curvature of the cavity dispersion relation and the fact that the light source, with a FWHM of 1 nm, is not narrow enough.}
\end{figure}

We performed measurements of Fourier rings by imaging the back focal plane of the collection lens onto a CMOS camera. 
Defects in the mirror coatings or a significant wedge between the mirrors will lead to deviations from a perfectly, circular ring.
Light from a laser diode with a FWHM of 1 nm was focused onto the cavity, and the transmitted light was collected from the Fourier plane. With this relatively narrow linewidth light source, constructive interference occurs only for specific in-plane k-vectors that satisfy the resonance condition, resulting in ring-shaped patterns on the camera. 
Figure \ref{fig:fourier_rings} shows these Fourier rings that were observed: in (a), a narrow ring with a physical diameter of approximately 11 mm is visible, whereas (b) shows a broader ring with a smaller diameter of approximately 4.1 mm. The imaging optics introduced a magnification factor of four between the Fourier plane and the CMOS sensor, corresponding to actual diameters of 2.8 mm and 1 mm in the Fourier plane for panels (a) and (b), respectively. These diameters correspond to in-plane momenta of $1.41\,\mu \text m^{-1}$ and $0.56\,\mu \text m^{-1}$.\\
The difference between (a) and (b) arises from an increase in the applied piezo-actuator voltage, which reduces the cavity length $L$. To maintain resonance with the fixed laser wavelength, one can see from Equation \ref{Eq:resonance_condition} that the angle $\theta$ has to decrease accordingly, leading to a smaller ring diameter. Additionally, because the laser has a finite spectral width of $\Delta \lambda$ several k-vectors can satisfy the resonance condition, particularly at smaller angles where the dispersion curve flattens (see Figure \ref{fig:cavity_disperion}). This results in broader ring features for smaller diameters, as observed in the experiment.

\newpage

\subsection{Cavity Stability} \label{Sec:stability}

To assess the vibrational behavior of the cavity, we used two photodiodes: one placed before the cavity to monitor the stability of the laser source, and another positioned after the cavity to measure the transmitted intensity. The cavity length was tuned such that the laser wavelength lay on the slope of a resonance peak, making the transmitted signal sensitive to small variations in cavity length.\\

\begin{figure}[htb!]
\includegraphics[width=\linewidth]{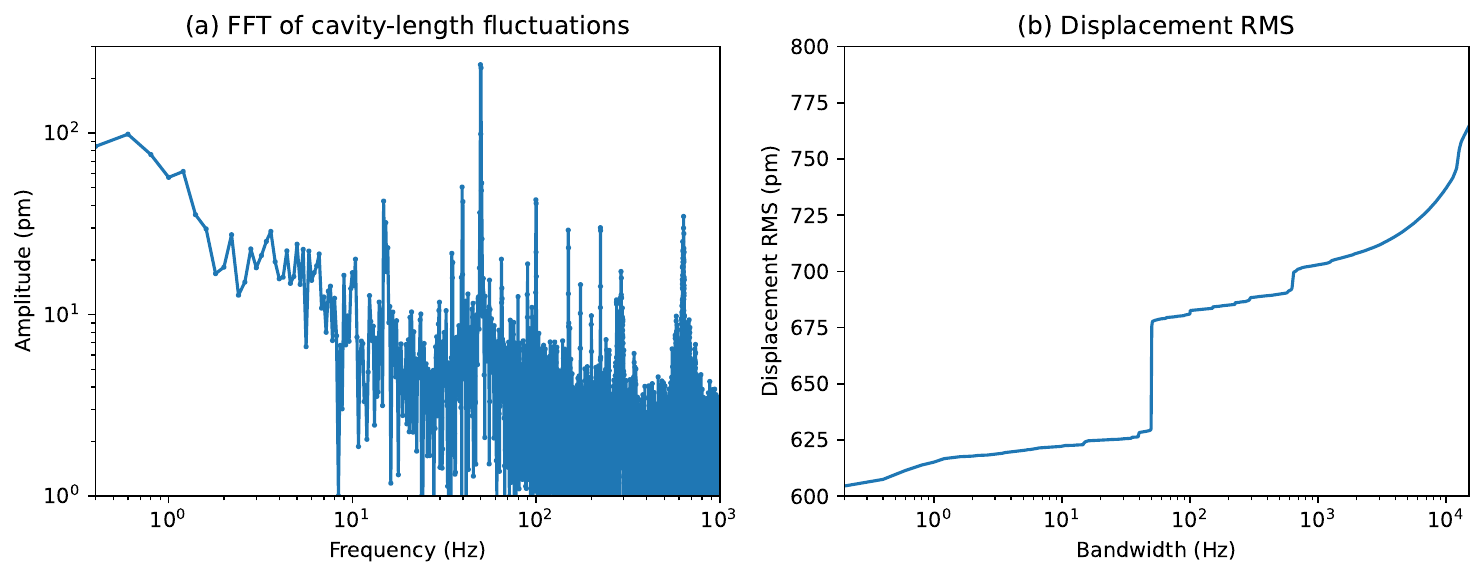}
\caption{\label{fig:cavity_stability} Vibrational analysis of the optical cavity: (a) FFT of cavity-length amplitude fluctuation (b) Integrated rms length fluctuations as function of bandwidth. The large step stems from residual 50 Hz line-frequency electrical noise coupling into the piezo voltage control.}
\end{figure}

With the photodiode in front of the cavity we confirmed the stability of the incident laser intensity throughout the experiment, indicating that any fluctuations in the transmitted signal originated from the cavity itself. We recorded the intensity of the transmitted signal as a function of time using the photodiode placed behind the cavity and performed a fast Fourier transform (FFT) of the amplified photo-current signal. Note that the laser operates on multiple longitudinal modes, which required a dedicated calibration procedure to extract the length to voltage conversion factor (see Appendix~\ref{Sec:stability_appendix} for details). Figure~\ref{fig:cavity_stability}(a) shows the FFT of the photodiode signal with a y-axis converted to length fluctuation amplitudes of the cavity length. We find that the cavity stability is better than 100 pm in a frequency band from 0.2 Hz to 10 kHz. Low frequency components up to approx. 1 Hz stem from slow drifts of the cavity. For higher frequencies above roughly 2 Hz the cavity stability is better than 50 pm. These stability levels are sufficient for operating the cavity in low-order longitudinal modes without active feedback. This showcases that our solid joint design is mechanically stable apart from the slow drifts that can be improved by optimizing the connection of mirror holder and piezo actuators. Figure~\ref{fig:cavity_stability}(b) shows the integrated rms displacement as function of bandwidth indicating that there are no sharp resonances in the frequency band up to 10 kHz. This indicates that the monolithic titanium frame and flexure-based coarse positioning do not introduce mechanical resonances in the relevant frequency band. The larger step of around 50 pm occurs at the mains frequency of 50 Hz (also visible as a peak in the FFT above 100 pm). This originates from residual line-frequency electrical noise coupling into the piezoelectric voltage control, which is subsequently translated into a mechanical cavity-length modulation and does not reflect a fundamental mechanical limitation of the cavity design.

Further improvement of the mechanical stability could be achieved by an optical \cite{Fisicaro2024} or electrical \cite{Yifan2018} measurement of the mirror separation. In order to avoid undesired modification of the sample (through local heating or optical doping) by large intracavity optical fields, electrical stabilization is preferable. To this end, at three points, ideally at the position of the three piezos, the mirror separation and tilt angle can be measured capacitively. The electrical readout signal can be used as a control signal for feedback onto the piezo voltage to stabilize the mirror separation at three points effectively stabilizing the cavity length and tilt angle.

\section{\label{sec:level1} Conclusion}
We presented a compact, open and mechanically stable 2D Fabry–Pérot microcavity that combines in situ tunability of mirror distance, wedge and cavity mode position. The symmetric optical design is optimized for angle-resolved transmission and reflection spectroscopy in real and momentum space and the platform is compatible with arbitrary (2D) materials. The access to k-space information simplifies data analysis and directly links the measured spectra to the in-plane cavity dispersion. While it does not reach the performance of monolithic cavities, our design addresses a complementary need: easy and fast integration of optically active materials into the cavity volume, rendering it a versatile platform for the study of more complex materials like vdW heterostructures. After placing the material of interest on the mirror substrate, it can be measured both inside and outside the cavity at any time. The open nature of the cavity assembly also allows for easy integration of complex structures like gated devices. For example, TMD layers can be embedded in an hBN-based van der Waals stack, which allows one to choose the thickness such that the TMD sits at an antinode and the gates at the node of the cavity field, while the in situ control of the mirror distance tunes the cavity mode across the exciton resonance. Graphene gates for electrical tuning of the exciton resonance can be contacted via wire bonds to pads on the non-etched mirror and routed out of the cavity making use of the mesa structure. Other emitting layers, such as perovskites, can be integrated in the same way. Furthermore, the in situ tunability of the mirror distance allows to control the detuning between optical transitions in the material and the cavity mode as well as the cavity length and thereby mode number. The ability to control the angle between the mirrors in situ allows to optimize the mirror parallelism as well as to introduce a well-defined wedge between the mirrors. In addition, it is possible to host an array of coupled lower dimensional cavity modes in the setup. This can be achieved by either patterning dielectrics onto the top mirror \cite{Shuangshuang2024} or by fabricating concave dimples into the top mirror before deposition of the mirror coating. This would enable generating a tunable array of coupled micro-cavities to study polaritonic quantum matter \cite{Basov2025}. In stark contrast to most other platforms the here presented design allows for transmission and thereby background free spectroscopy. This is of particular importance for studying polariton interactions using photon correlation spectroscopy. While presently we focused on room-temperature operation, the cavity assembly is kept as compact as possible to allow for integration into an optical closed-cycle cryostat. The achieved finesse of 360, Q=3750, and cavity-length fluctuations below 100 pm establish our system as a tool for room-temperature cavity spectroscopy and a basis for future cryogenic or strong-coupling experiments.

\appendix
\section{Mesa fabrication process}
The mesa structure on the fused silica substrate is fabricated using a chemical etching process based on highly concentrated hydrogen fluoride. With an acid concentration of 40\%, we observed etch rates of $0.7\,\mu \textrm{m/min}$. Hence, a significant step of around $50 \mu m$ between the mirror surface and the substrate requires a lithographic mask to withstand the etch for about one hour. We achieve this by a metallization of 30 nm of chrome for better adhesion and 400 nm of gold. We then apply a negative photoresist (AZ nLOF 2070) and expose circular areas of $400\,\mu m$ diameter. After the development, we baked the resist for 30 min to improve adhesion and harden it, such that it acts as an additional protection layer of several micrometer thickness. The rest of the metal is then removed before the final etching process with hydrofluoric acid is applied for about one hour. Due to the isotropic etching of hydrofluoric acid, the mask is under-etched, reducing the effective diameter to around $250\mu$m. During the process, we observed that a thorough cleaning procedure before chrome application was vital to reduce the lateral underetching and significantly decrease the number of pits on the final mesa.

\section{Real space image of cavity transmission mode}

\begin{figure}[htb!]
    \centering
    \includegraphics[width=0.48\linewidth]{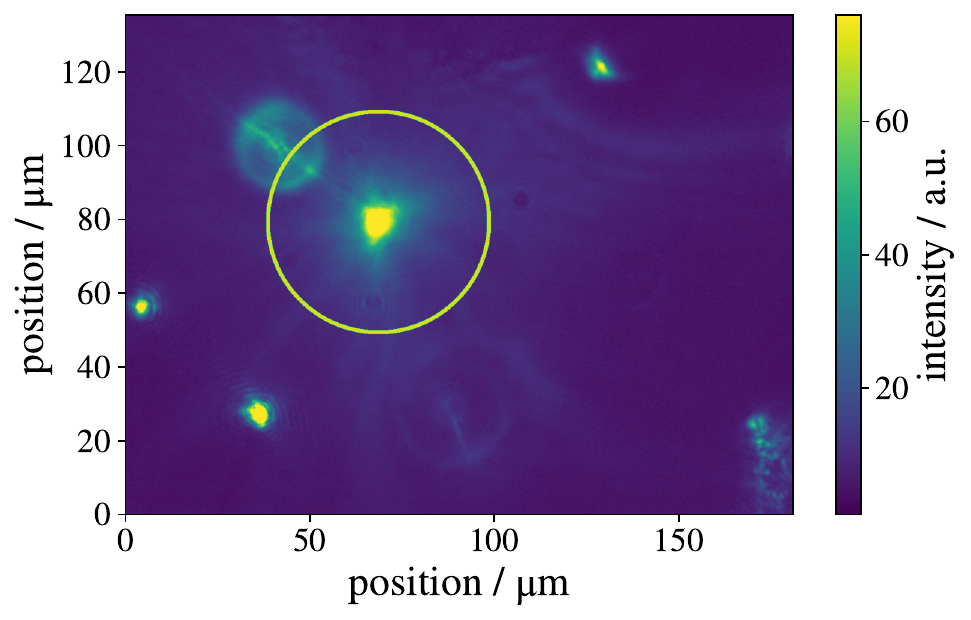}
\caption{\label{fig:detection_mode_on_mesa} Imaging the cavity mode on the mesa structure: The circle in the picture represents the aperture of 60 $\mu$m from which the transmitted light of the cavity mode can be detected. }
\end{figure}

In Figure \ref{fig:detection_mode_on_mesa} we show the image of the transmitted cavity mode on the backside of the mesa. The measurement was performed in the configuration where the light was focused onto the mesa. The transmitted mode is at the position x=75$\mu$m and y=80$\mu$m. Apart from the transmitted mode there are several bright regions where light exits the mirror at defect positions of the mirror (round shapes). The region around x=175$\mu$m and y=10$\mu$m is a region where scattered light exits the edge of the etched mesa walls. The yellow circle around the transmitted mode marks the approximate area of the detection mode when measured in the configuration of a k-resolved detection. Since this area is mostly free from defects, we measure the smallest linewidths here.

\newpage

\section{Cavity stability measurement methods} \label{Sec:stability_appendix}

The cavity-length stability is extracted from fluctuations of the transmitted photodiode signal with the cavity operated on the linear slope of a transmission resonance. In this regime, small variations of the cavity length result in proportional changes of the detected transmission. A quantitative conversion of the measured photodiode signal into an equivalent cavity-length fluctuation therefore requires a calibration of the intensity-to-length conversion factor.

\begin{figure}[htb!]
    \subfigure[]{\includegraphics[width=0.48\linewidth]{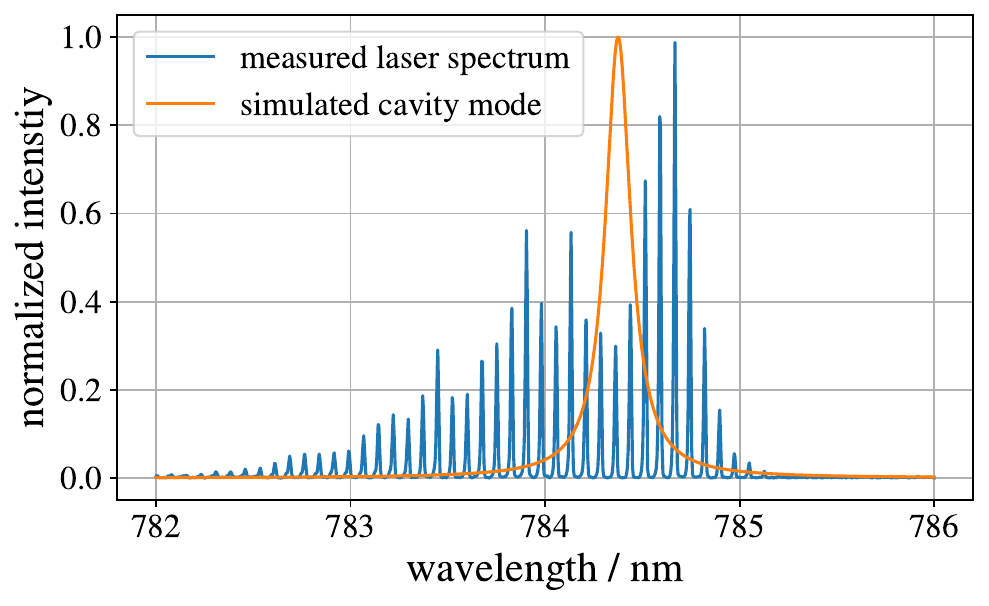}}
    \subfigure[]{\includegraphics[width=0.48\linewidth]{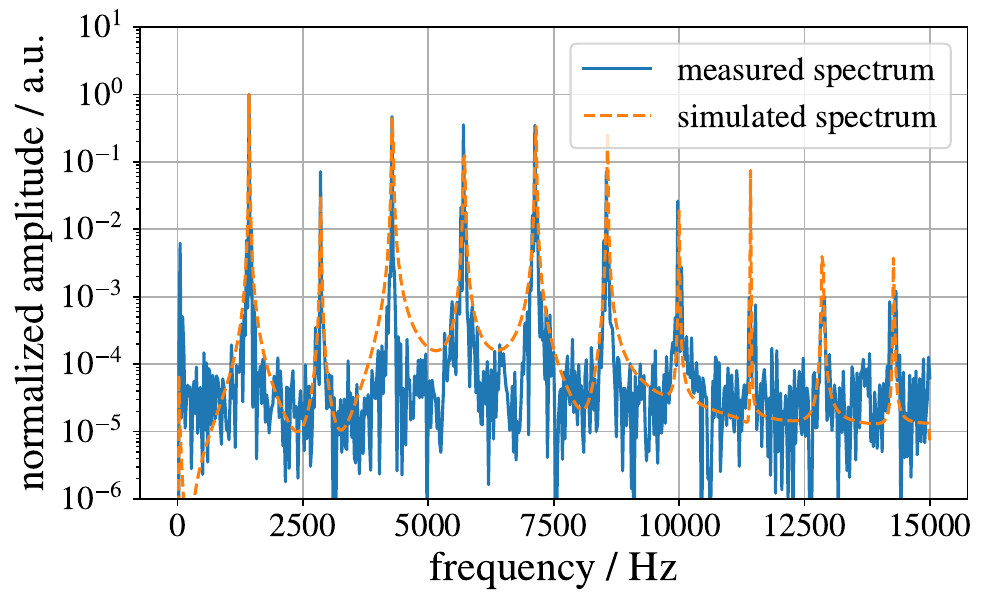}}
\caption{\label{fig:cavity_stability_simulation} Simulated cavity stability measurement: (a) shows the spectrum of the laser diode and an ideal cavity mode with FWHM of 0.2 nm, in (b) we show the comparison between the simulated power spectrum of a cavity mode centered at 784.4 nm vibrating at a frequency of 1427 Hz with an amplitude of 2.2 nm and our experimental measurements.}
\end{figure}

Ideally, the measurement in Section \ref{Sec:stability} would be performed with a single-mode laser featuring a linewidth much narrower than the cavity mode. However, our laser has a 1 nm FWHM and has multiple longitudinal modes, as can be seen in Figure \ref{fig:cavity_stability_simulation} (a), resulting in a convolution of the laser and cavity modes in the observed signal.\\

\begin{figure}[htb!]
    \subfigure[]{\includegraphics[width=0.48\linewidth]{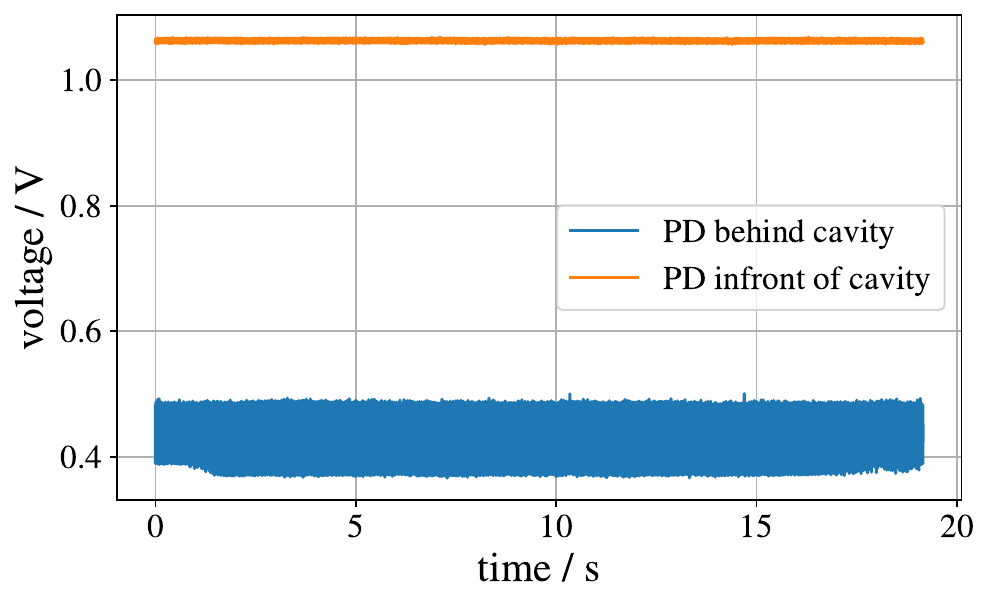}}
    \subfigure[]{\includegraphics[width=0.48\linewidth]{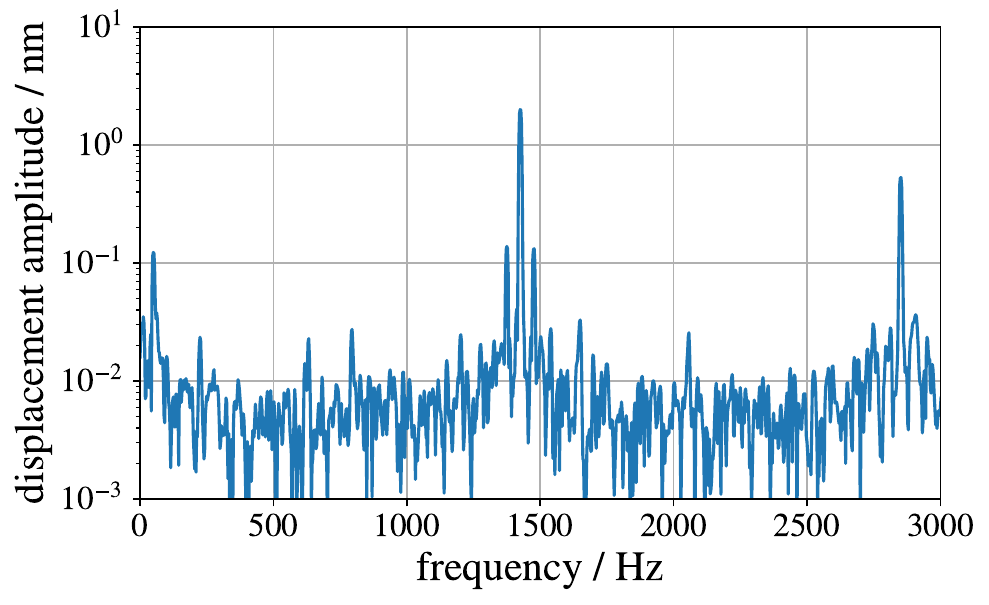}}
\caption{\label{fig:cavity_stability_with_modulation} Vibrational analysis of the optical cavity: (a) shows the time signal of two photodiodes, which monitor the intensity of the excitation light, in (b) the power spectrum of the signal from the PD behind the cavity is shown. One can observe a prominent peak at 1427 Hz which is the modulation frequency.}
\end{figure}

To quantify the vibration amplitude, we applied a sinusoidal modulation with 1427 Hz to the piezoelectric actuators in a separate measurement, introducing a controlled 2 nm variation in cavity length. The resulting transmitted intensity and its Fourier spectrum revealed a prominent peak at the modulation frequency, along with higher-frequency components due to the additional modes of the laser, Figure \ref{fig:cavity_stability_simulation} (b). Therefore, the amplitude of the 2 nm modulation was distributed across several frequencies. For the calibration of the displacement amplitude (y-axis) of our Fourier spectra we assumed that the entire 2 nm displacement contributed to the peak at 1427 Hz. This defines the y-axis scale in Figure \ref{fig:cavity_stability_with_modulation} (b) in case of a modulated cavity. This scale is then used in the Fourier spectrum without modulation, Figure \ref{fig:cavity_stability} (a) and provides a lower bound on the cavity stability. The true cavity stability is even better than this. 

\newpage

\section*{Funding}
ZHAW Zurich University of Applied Sciences.

\section*{Data Availability Statement}

The data that support the findings of this study are available from the
corresponding author upon reasonable request.

\section*{Disclosures}

The authors declare that there are no conflicts of interest related to this article.

\bibliography{cavity}

@Article{Shang2023,
author={Shang, Jingzhi
and Zhang, Xinyu
and Zhang, Vanessa Li
and Zhang, Xuewen
and Yu, Ting},
title={{Exciton--Photon Interactions in Two-Dimensional Semiconductor Microcavities}},
journal={ACS Photonics},
year={2023},
month={Jul},
day={19},
publisher={American Chemical Society},
volume={10},
number={7},
pages={2064-2077},
doi={10.1021/acsphotonics.2c01541},
url={https://doi.org/10.1021/acsphotonics.2c01541}
}

@Article{Kasprzak2006,
author={Kasprzak, J.
and Richard, M.
and Kundermann, S.
and Baas, A.
and Jeambrun, P.
and Keeling, J. M. J.
and Marchetti, F. M.
and Szyma{\'{n}}ska, M. H.
and Andr{\'e}, R.
and Staehli, J. L.
and Savona, V.
and Littlewood, P. B.
and Deveaud, B.
and Dang, Le Si},
title={{Bose--Einstein condensation of exciton polaritons}},
journal={Nature},
year={2006},
month={Sep},
day={01},
volume={443},
number={7110},
pages={409-414},
issn={1476-4687},
doi={10.1038/nature05131},
url={https://doi.org/10.1038/nature05131}
}

@article{carusotto2013,
  title={{Quantum fluids of light}},
  author={Carusotto, Iacopo and Ciuti, Cristiano},
  journal={Reviews of Modern Physics},
  volume={85},
  number={1},
  pages={299--366},
  year={2013},
  publisher={APS}
}

@Article{Amo2009,
author={Amo, Alberto
and Lefr{\`e}re, J{\'e}r{\^o}me
and Pigeon, Simon
and Adrados, Claire
and Ciuti, Cristiano
and Carusotto, Iacopo
and Houdr{\'e}, Romuald
and Giacobino, Elisabeth
and Bramati, Alberto},
title={{Superfluidity of polaritons in semiconductor microcavities}},
journal={Nature Physics},
year={2009},
month={Nov},
day={01},
volume={5},
number={11},
pages={805-810},
issn={1745-2481},
doi={10.1038/nphys1364},
url={https://doi.org/10.1038/nphys1364}
}

@Article{Nardin2011,
author={Nardin, Ga{\"e}l
and Grosso, Gabriele
and L{\'e}ger, Yoan
and Piȩtka, Barbara
and Morier-Genoud, Fran{\c{c}}ois
and Deveaud-Pl{\'e}dran, Beno{\^i}t},
title={{Hydrodynamic nucleation of quantized vortex pairs in a polariton quantum fluid}},
journal={Nature Physics},
year={2011},
month={Aug},
day={01},
volume={7},
number={8},
pages={635-641},
issn={1745-2481},
doi={10.1038/nphys1959},
url={https://doi.org/10.1038/nphys1959}
}

@Article{Schneider2013,
author={Schneider, Christian
and Rahimi-Iman, Arash
and Kim, Na Young
and Fischer, Julian
and Savenko, Ivan G.
and Amthor, Matthias
and Lermer, Matthias
and Wolf, Adriana
and Worschech, Lukas
and Kulakovskii, Vladimir D.
and Shelykh, Ivan A.
and Kamp, Martin
and Reitzenstein, Stephan
and Forchel, Alfred
and Yamamoto, Yoshihisa
and H{\"o}fling, Sven},
title={{An electrically pumped polariton laser}},
journal={Nature},
year={2013},
month={May},
day={01},
volume={497},
number={7449},
pages={348-352},
issn={1476-4687},
doi={10.1038/nature12036},
url={https://doi.org/10.1038/nature12036}
}

@article{Chervy2020,
  title = {{Accelerating Polaritons with External Electric and Magnetic Fields}},
  author = {Chervy, T. and Kn\"uppel, P. and Abbaspour, H. and Lupatini, M. and F\"alt, S. and Wegscheider, W. and Kroner, M. and Imamo\ifmmode \check{g}\else \v{g}\fi{}lu, A.},
  journal = {Phys. Rev. X},
  volume = {10},
  issue = {1},
  pages = {011040},
  numpages = {7},
  year = {2020},
  month = {Feb},
  publisher = {American Physical Society},
  doi = {10.1103/PhysRevX.10.011040},
  url = {https://link.aps.org/doi/10.1103/PhysRevX.10.011040}
}

@Article{Mak2016,
author={Mak, Kin Fai
and Shan, Jie},
title={{Photonics and optoelectronics of 2D semiconductor transition metal dichalcogenides}},
journal={Nature Photonics},
year={2016},
month={Apr},
day={01},
volume={10},
number={4},
pages={216-226},
issn={1749-4893},
doi={10.1038/nphoton.2015.282},
url={https://doi.org/10.1038/nphoton.2015.282}
}

@Article{Manser2016,
author={Manser, Joseph S.
and Christians, Jeffrey A.
and Kamat, Prashant V.},
title={{Intriguing Optoelectronic Properties of Metal Halide Perovskites}},
journal={Chemical Reviews},
year={2016},
month={Nov},
day={09},
publisher={American Chemical Society},
volume={116},
number={21},
pages={12956-13008},
issn={0009-2665},
doi={10.1021/acs.chemrev.6b00136},
url={https://doi.org/10.1021/acs.chemrev.6b00136}
}

@article{Rupprecht2021,
    author = {Rupprecht, Christoph and Lundt, Nils and Wurdack, Matthias and Stepanov, Petr and Estrecho, Eliezer and Richard, Maxime and Ostrovskaya, Elena A. and Höfling, Sven and Schneider, Christian},
    title = {{Micro-mechanical assembly and characterization of high-quality Fabry–Pérot microcavities for the integration of two-dimensional materials}},
    journal = {Applied Physics Letters},
    volume = {118},
    number = {10},
    pages = {103103},
    year = {2021},
    month = {03},
    issn = {0003-6951},
    doi = {10.1063/5.0034851},
    url = {https://doi.org/10.1063/5.0034851},
}

@Article{Dufferwiel2015,
author={Dufferwiel, S.
and Schwarz, S.
and Withers, F.
and Trichet, A. A. P.
and Li, F.
and Sich, M.
and Del Pozo-Zamudio, O.
and Clark, C.
and Nalitov, A.
and Solnyshkov, D. D.
and Malpuech, G.
and Novoselov, K. S.
and Smith, J. M.
and Skolnick, M. S.
and Krizhanovskii, D. N.
and Tartakovskii, A. I.},
title={{Exciton--polaritons in van der Waals heterostructures embedded in tunable microcavities}},
journal={Nature Communications},
year={2015},
month={Oct},
day={08},
volume={6},
number={1},
pages={8579},
issn={2041-1723},
doi={10.1038/ncomms9579},
url={https://doi.org/10.1038/ncomms9579}
}

@Article{Sidler2017,
author={Sidler, Meinrad
and Back, Patrick
and Cotlet, Ovidiu
and Srivastava, Ajit
and Fink, Thomas
and Kroner, Martin
and Demler, Eugene
and Imamoglu, Atac},
title={{Fermi polaron-polaritons in charge-tunable atomically thin semiconductors}},
journal={Nature Physics},
year={2017},
month={Mar},
day={01},
volume={13},
number={3},
pages={255-261},
issn={1745-2481},
doi={10.1038/nphys3949},
url={https://doi.org/10.1038/nphys3949}
}

@article{Vadia2021,
  title = {{Open-Cavity in Closed-Cycle Cryostat as a Quantum Optics Platform}},
  author = {Vadia, Samarth and Scherzer, Johannes and Thierschmann, Holger and Sch\"afermeier, Clemens and Dal Savio, Claudio and Taniguchi, Takashi and Watanabe, Kenji and Hunger, David and Karra\"{\i}, Khaled and H\"ogele, Alexander},
  journal = {PRX Quantum},
  volume = {2},
  issue = {4},
  pages = {040318},
  numpages = {10},
  year = {2021},
  month = {Oct},
  publisher = {American Physical Society},
  doi = {10.1103/PRXQuantum.2.040318},
  url = {https://link.aps.org/doi/10.1103/PRXQuantum.2.040318}
}

@Article{Drawer2023,
author={Drawer, Jens-Christian
and Mitryakhin, Victor Nikolaevich
and Shan, Hangyong
and Stephan, Sven
and Gittinger, Moritz
and Lackner, Lukas
and Han, Bo
and Leibeling, Gilbert
and Eilenberger, Falk
and Banerjee, Rounak
and Tongay, Sefaattin
and Watanabe, Kenji
and Taniguchi, Takashi
and Lienau, Christoph
and Silies, Martin
and Anton-Solanas, Carlos
and Esmann, Martin
and Schneider, Christian},
title={{Monolayer-Based Single-Photon Source in a Liquid-Helium-Free Open Cavity Featuring 65{\%} Brightness and Quantum Coherence}},
journal={Nano Letters},
year={2023},
month={Sep},
day={27},
publisher={American Chemical Society},
volume={23},
number={18},
pages={8683-8689},
issn={1530-6984},
doi={10.1021/acs.nanolett.3c02584},
url={https://doi.org/10.1021/acs.nanolett.3c02584}
}

@PHDTHESIS{Wuester2015,
	copyright = {In Copyright - Non-Commercial Use Permitted},
	year = {2015},
	type = {Doctoral Thesis},
	author = {Wüster, Wolf},
	size = {1 Band},
	address = {Zurich},
	publisher = {ETH Zurich},
	DOI = {10.3929/ethz-a-010542649},
	title = {{Cavity quantum electrodynamics with many-body states of a two-dimensional electron system}},
	school = {ETH Zurich}
}

@article{Skolnick1998,
doi = {10.1088/0268-1242/13/7/003},
url = {https://dx.doi.org/10.1088/0268-1242/13/7/003},
year = {1998},
month = {jul},
publisher = {},
volume = {13},
number = {7},
pages = {645},
author = {M S Skolnick and T A Fisher and D M Whittaker},
title = {{Strong coupling phenomena in quantum microcavity structures}},
journal = {Semiconductor Science and Technology}
}

@article{Brovelli1995,
title = {{Simple analytical expressions for the reflectivity and the penetration depth of a Bragg mirror between arbitrary media}},
journal = {Optics Communications},
volume = {116},
number = {4},
pages = {343-350},
year = {1995},
issn = {0030-4018},
doi = {https://doi.org/10.1016/0030-4018(95)00084-L},
url = {https://www.sciencedirect.com/science/article/pii/003040189500084L},
author = {L.R Brovelli and U Keller}
}

@article{Knopf2019,
author = {Heiko Knopf and Nils Lundt and Tobias Bucher and Sven H\"{o}fling and Sefaattin Tongay and Takashi Taniguchi and Kenji Watanabe and Isabelle Staude and Ulrike Schulz and Christian Schneider and Falk Eilenberger},
journal = {Opt. Mater. Express},
number = {2},
pages = {598--610},
publisher = {Optica Publishing Group},
title = {{Integration of atomically thin layers of transition metal dichalcogenides into high-Q, monolithic Bragg-cavities: an experimental platform for the enhancement of the optical interaction in 2D-materials}},
volume = {9},
month = {Feb},
year = {2019},
url = {https://opg.optica.org/ome/abstract.cfm?URI=ome-9-2-598},
doi = {10.1364/OME.9.000598},
}

@Article{Lopriore2025,
author={Lopriore, Edoardo
and Tagarelli, Fedele
and Fitzgerald, Jamie M.
and Gonzalez Marin, Juan Francisco
and Watanabe, Kenji
and Taniguchi, Takashi
and Malic, Ermin
and Kis, Andras},
title={{Enhancing interlayer exciton dynamics by coupling with monolithic cavities via the field-induced Stark effect}},
journal={Nature Nanotechnology},
year={2025},
month={Oct},
day={01},
volume={20},
number={10},
pages={1412-1418},
issn={1748-3395},
doi={10.1038/s41565-025-01969-2},
url={https://doi.org/10.1038/s41565-025-01969-2}
}

@article{Hunger_2010,
doi = {10.1088/1367-2630/12/6/065038},
url = {https://doi.org/10.1088/1367-2630/12/6/065038},
year = {2010},
month = {Jun},
publisher = {},
volume = {12},
number = {6},
pages = {065038},
author = {Hunger, D and Steinmetz, T and Colombe, Y and Deutsch, C and Hänsch, T W and Reichel, J},
title = {{A fiber Fabry–Perot cavity with high finesse}},
journal = {New Journal of Physics}
}

@article{Lackner2021,
author={Lackner, L.
and Dusel, M.
and Egorov, O. A.
and Han, B.
and Knopf, H.
and Eilenberger, F.
and Schr{\"o}der, S.
and Watanabe, K.
and Taniguchi, T.
and Tongay, S.
and Anton-Solanas, C.
and H{\"o}fling, S.
and Schneider, C.},
title={{Tunable exciton-polaritons emerging from WS2 monolayer excitons in a photonic lattice at room temperature}},
journal={Nature Communications},
year={2021},
month={Aug},
day={16},
volume={12},
number={1},
pages={4933},
issn={2041-1723},
doi={10.1038/s41467-021-24925-9},
url={https://doi.org/10.1038/s41467-021-24925-9}
}

@article{Lackner2025,
  title = {{Topologically Tunable Polaritons Based on a Two-Dimensional Crystal in a Photonic Lattice}},
  author = {Lackner, L. and Egorov, O. A. and Ernzerhof, A. and Bennenhei, C. and Mitryakhin, V. N. and Leibeling, G. and Eilenberger, F. and Tongay, S. and Peschel, U. and Esmann, M. and Schneider, C.},
  journal = {Phys. Rev. Lett.},
  volume = {135},
  issue = {16},
  pages = {166901},
  numpages = {6},
  year = {2025},
  month = {Oct},
  publisher = {American Physical Society},
  doi = {10.1103/qphl-qd37},
  url = {https://link.aps.org/doi/10.1103/qphl-qd37}
}

@article{Krol2023,
author = {M. Kr\'{o}l and K. {\L}empicka-Mirek and K. Rechci\'{n}ska and M. Furman and K. Nogajewski and R. Mazur and P. Morawiak and W. Piecek and W. Pacuski and J. Szczytko and B. Piętka},
journal = {Opt. Mater. Express},
number = {9},
pages = {2651--2661},
publisher = {Optica Publishing Group},
title = {{Universality of open microcavities for strong light-matter coupling}},
volume = {13},
month = {Sep},
year = {2023},
url = {https://opg.optica.org/ome/abstract.cfm?URI=ome-13-9-2651},
doi = {10.1364/OME.492285},
}

@article{Krol2020,
doi = {10.1088/2053-1583/ab4b14},
url = {https://doi.org/10.1088/2053-1583/ab4b14},
year = {2019},
month = {Oct},
publisher = {IOP Publishing},
volume = {7},
number = {1},
pages = {015006},
author = {Król, Mateusz and Rechcińska, Katarzyna and Nogajewski, Karol and Grzeszczyk, Magdalena and Łempicka, Karolina and Mirek, Rafał and Piotrowska, Sara and Watanabe, Kenji and Taniguchi, Takashi and Molas, Maciej R and Potemski, Marek and Szczytko, Jacek and Piętka, Barbara},
title = {{Exciton-polaritons in multilayer WSe2 in a planar microcavity}},
journal = {2D Materials}
}

@article{Levinsen2019,
  title = {{Spectroscopic Signatures of Quantum Many-Body Correlations in Polariton Microcavities}},
  author = {Levinsen, Jesper and Marchetti, Francesca Maria and Keeling, Jonathan and Parish, Meera M.},
  journal = {Phys. Rev. Lett.},
  volume = {123},
  issue = {26},
  pages = {266401},
  numpages = {6},
  year = {2019},
  month = {Dec},
  publisher = {American Physical Society},
  doi = {10.1103/PhysRevLett.123.266401},
  url = {https://link.aps.org/doi/10.1103/PhysRevLett.123.266401}
}

@Article{Delteil2019,
author={Delteil, Aymeric
and Fink, Thomas
and Schade, Anne
and H{\"o}fling, Sven
and Schneider, Christian
and {\.{I}}mamo{\u{g}}lu, Ata{\c{c}}},
title={{Towards polariton blockade of confined exciton--polaritons}},
journal={Nature Materials},
year={2019},
month={Mar},
day={01},
volume={18},
number={3},
pages={219-222},
issn={1476-4660},
doi={10.1038/s41563-019-0282-y},
url={https://doi.org/10.1038/s41563-019-0282-y}
}

@Article{Scarpelli2024,
author={Scarpelli, Lorenzo
and Elouard, Cyril
and Johnsson, Mattias
and Morassi, Martina
and Lemaitre, Aristide
and Carusotto, Iacopo
and Bloch, Jacqueline
and Ravets, Sylvain
and Richard, Maxime
and Volz, Thomas},
title={{Probing many-body correlations using quantum-cascade correlation spectroscopy}},
journal={Nature Physics},
year={2024},
month={Feb},
day={01},
volume={20},
number={2},
pages={214-218},
issn={1745-2481},
doi={10.1038/s41567-023-02322-x},
url={https://doi.org/10.1038/s41567-023-02322-x}
}

@misc{Byrnes2016,
      title={{Multilayer optical calculations}}, 
      author={Steven J. Byrnes},
      year={2020},
      eprint={1603.02720},
      archivePrefix={arXiv},
      primaryClass={physics.comp-ph},
      url={https://arxiv.org/abs/1603.02720},
      note={{arXiv:1603.02720v5}},
}

@Article{Yifan2018,
author = {Yifan Lu and Zhijian Cai and Peiliang Guo and Yanfeng Su and Jianhong Wu},
title = {{An automatic cavity-length stabilizing and monitoring system for tunable Fabry-Perot filter}},
volume = {10827},
booktitle = {Sixth International Conference on Optical and Photonic Engineering (icOPEN 2018)},
editor = {Yingjie Yu and Chao Zuo and Kemao Qian},
journal = {Proc. SPIE 10827, Sixth International Conference on Optical and Photonic Engineering},
organization = {International Society for Optics and Photonics},
publisher = {SPIE},
pages = {108270I},
year = {2018},
doi = {10.1117/12.2501033},
URL = {https://doi.org/10.1117/12.2501033}
}

@article{Fisicaro2024,
    author = {Fisicaro, M. and Witlox, M. and van der Meer, H. and Löffler, W.},
    title = {Active stabilization of an open-access optical microcavity for low-noise operation in a standard closed-cycle cryostat},
    journal = {Review of Scientific Instruments},
    volume = {95},
    number = {3},
    pages = {033101},
    year = {2024},
    month = {03},
    issn = {0034-6748},
    doi = {10.1063/5.0174982},
    url = {https://doi.org/10.1063/5.0174982},
    eprint = {https://pubs.aip.org/aip/rsi/article-pdf/doi/10.1063/5.0174982/19704946/033101_1_5.0174982.pdf},
}

@article{hoang2026,
  title={A compact, robust, and tunable open microcavity platform for solid-state quantum electrodynamics},
  author={Hoang, Thi D and Mahdikhany, Fateme and Wang, Zixuan and Mirin, Richard and Silverman, Kevin and Imany, Poolad and Sun, Shuo},
  journal={Optica Quantum},
  volume={4},
  number={4},
  pages={359--365},
  year={2026},
  publisher={Optica Publishing Group}
}

@article{Shuangshuang2024,
    author = {Tian, Shuangshuang and Wang, Qi and Liang, Shuang and Han, Qi and Zhang, Debao and Huang, Zhongmin and Ning, Jiqiang and Mei, Shiliang and Xie, Wei and Zhao, Haibin and Wu, Xiang and Wang, Jun},
    title = {High Q‑Factor Single-Mode
Lasing in Inorganic Perovskite Microcavities with Microfocusing Field
Confinement},
    journal = {Nano Letters},
    volume = {24},
    number = {4},
    pages = {1406-1414},
    year = {2024},
    month = {01},
    issn = {1530-6984},
    doi = {10.1021/acs.nanolett.3c04797},
    url = {https://doi.org/10.1021/acs.nanolett.3c04797},
    eprint = {https://pubs.acs.org/nalefd/article-pdf/24/4/1406/9551937/nl3c04797.pdf},
}

@article{Basov2025,
url = {https://doi.org/10.1515/nanoph-2025-0001},
title = {Polaritonic quantum matter},
title = {},
author = {D.N. Basov and Ana Asenjo-Garcia and P. James Schuck and Xiaoyang Zhu and Angel Rubio and Andrea Cavalleri and Milan Delor and Michael M. Fogler and Mengkun Liu},
pages = {3723--3760},
volume = {14},
number = {23},
journal = {Nanophotonics},
doi = {doi:10.1515/nanoph-2025-0001},
year = {2025},
lastchecked = {2026-11-02}
}

\end{document}